\input harvmac.tex
\vskip 2in
\Title{\vbox{\baselineskip12pt
\hbox to \hsize{\hfill}
\hbox to \hsize{\hfill}}}
{\vbox{\centerline{ GL(1) Charged States in Twistor String Theory }
\vskip 0.3in
{\vbox{\centerline{}}}}}
\centerline{Dimitri Polyakov\footnote{$^\dagger$}
{dp02@aub.edu.lb}}
\medskip
\centerline{\it Center for Advanced Mathematical Studies}
\centerline{\it and  Department of Physics }
\centerline{\it  American University of Beirut}
\centerline{\it Beirut, Lebanon}
\vskip .5in
\centerline {\bf Abstract}
We discuss the appearance of the GL(1) charged physical operators
in the twistor string theory. These operators
are shown to be BRST-invariant and non-trivial and some of their
correlators and conformal $\beta$-functions are computed.
Remarkably, the non-conservation of the $GL(1)$ charge
in interactions involving these operators, is related to the anomalous
term in the Kac-Moody current algebra.
While these operators play no role in the maximum helicity violating
(MHV) amplitudes, they are shown to contribute nontrivially to the non-MHV
correlators in the presence of the worldsheet instantons.
We argue that these operators describe the non-perturbative dynamics of
solitons in conformal supergravity. The exact form of such solitonic solutions
is yet to be determined.
 {\bf}
\Date{December 2004}
\vfill\eject
\lref\pol{ S. Gubser, I. Klebanov, A.M. Polyakov, Phys. Lett. B428, 105 (1998)}
\lref\wit{E. Witten, Adv. Theor. Math. Phys, 2, 253 (1998)}
\lref\malda{J. Maldacena, Adv. Theor. Math. Phys. 2, 231 (1998)}
\lref\witten{E. Witten,hep-th/0312171, Commun. Math. Phys.252:189(2004)}
\lref\wb{N. Berkovits, E. Witten, hep-th/0406051, JHEP 0408:009(2004)}
\lref\cw{F. Cachazo, P. Svrcek, E. Witten, hep-th/0406177, JHEP 0410:074
 (2004)}
\lref\mb{N. Berkovits, L. Motl, hep-th/0403187, JHEP 0404:056 (2004)}
\lref\self{D. Polyakov, hep-th/0406079, to appear in IJMPA}
\lref\witt{ E. Witten, hep-th/9112056}
\lref\town{J.Azcarraga, J.Gauntlett, J.Izquierdo, P.Townsend, Ph.Rev.Lett.
D63(1989)2443}
\centerline{\bf Introduction}
The hypothesis of  the gauge-string correspondence, attempting
to relate the gauge-theoretic and string degrees of freedom,
 is a long-standing problem 
of great importance. Remarkably, such a correspondence
can be shown to occur both in the strongly coupled and 
the perturbative limits of Yang-Mills theory.
Firstly, it is well-known that the AdS/CFT correspondence
(e.g. see {\pol, \malda, \wit}) implies the
isomorphism between vertex operators of string theory in the anti de Sitter 
backgrounds and the local gauge-invariant observables on the Yang-Mills side.
On the other hand, recently it has been realized that some of the perturbative
Yang-Mills  scattering amplitudes are reproduced by the correlators
of string theory with the twistorial target space and the related
topological B-model with Calabi-Yau target space {\witt}
The open string amplitudes of the twistor-string theory, supported on 
holomorphic curves in the projective space, have been shown
to reproduce the perturbative gauge-theoretic ampludes and the 
perturbative expansion on the Yang-Mills side have been shown 
to correspond to the D-instanton (or the worldsheet instanton) expansion
in string theory {\witten, \cw}.
 Typically, the worldsheet correlators 
of twistorial string theory involve the GL(1)-neutral vertex operators
from both open and closed sectors. As already has been noted,
open string amplitudes
reproduce the Yang-Mills perturbation theory, while
the closed string modes describe the conformal supergravity in the 
low-energy limit {\wb, \mb}
In this letter, we show that, apart from the usual GL(1)-neutral operators,
the twistor string theory also contains a set of $physical$ 
GL(1)-charged vertex operators, argued to be related
to the non-perturbative dynamics of the solitons appearing in 
conformal supergravity. 
These operators can be thought of as the second-quantized creation
 operators for the solitonic
solutions of the conformal supergravity.
Unfortunately, the precise form of these solutions in terms of the
conformal supergravity fields remains undetermined in this paper and
this is left for the future work.
 While these new vertex operators  
play no role in the MHV
amplitudes, their role becomes important in the presence of the D-instantons
(or, equivalently, in the non-MHV cases)
and they do contribute nontrivially
 to both open and closed string amplitudes of the twistorial string theory.
In this letter, we construct  few examples of such GL(1)-charged BRST-invariant
operators and derive the 
conditions for their BRST-invariance and non-triviality.
At the first glance, appearance of such new physical vertices
in twistor-string theory presents a problem as these states would
seem to be in contradiction with
the GL(1) gauge symmetry of the theory.
One should remember, however, that in string theory BRST-invariance
of some physical vertex operator does not automatically imply its 
gauge invariance. The simplest example is the local worldsheet supersymmetry
in the RNS superstring model; e.g.
 the unintegrated picture $-1$ vertex operator of a photon
is of course BRST-invariant; but straightforvard
application of the worldsheet supersymmetry generator to this vertex
shows its non-invariance under the local supersymmetry.
Despite such a  non-invariance, however, the relevant scattering amplitudes 
 $are$ invariant under the local supersymmetry, provided that one fixes
the gauge parameter to vanish at the insertion points of the vertices.
Such constraints on the gauge transformation parameter effectively reduces
the gauge group, leading to appearance of the picture-changing oprators in the
amplitudes. The situation is somewhat similar in the twistor string theory,
even though the complete  and well-defined picture-changing formalism 
is yet to be  developed for twistorial strings.

 While the main purpose of this letter is to point out the existence 
the $GL(1)$-charged physical operators and to discuss  some of their properties
and amplitudes, we also shall attempt
(mainly by using the NSR superstring analogies) 
to present some physical arguments
relating these states to the solitons of conformal supergravity.
This will be left to the concluding discussion section of this letter.

\centerline{\bf $GL(1)$-charged operators and their amplitudes}

The worldsheet action for the twistor string theory
in conformal gauge is given by
\eqn\grav{\eqalign{S=\int{d^2z}(Y_{I}{{\nabla}_{\bar{z}}}Z^{I}
+{\bar{Y}}_{I}{\nabla_z}{{\bar{Z}}^{I}})+S_{c}\cr
\nabla_z=\partial-A_z;\nabla_{\bar{z}}=\bar\partial-A_{\bar{z}}
}}

where the supertwistors
$Z^I=(\lambda^{\alpha},\mu^{{\dot{\alpha}}},\psi^A)$
are homogeneous coordinates in the $CP^{3|4}$ target space,
$Y_I=({\bar\mu}_\alpha,{\bar\lambda}_{{\dot{\alpha}}},{\bar\psi}_A)$
are conjugate to $Z$, $A_a(a=1,2)$ is the $GL(1)$ connection 
and $S_c$ is the action for a system
of currents with $c=28$, necessary to cancel the conformal anomaly.
In the open string case, considered in this paper,
$CP^{3|4}$ is reduced to $RP^{3|4}$ and $Z^I=\bar{Z^I},Y^I=\bar{Y^I}$
on the worldsheet boundary.
When worldsheet instantons are absent,
the stress-energy tensor and the $GL(1)$-generator are given by
\eqn\grav{\eqalign{T=Y_I{\partial}{Z^I}+T_c\cr
J_{GL(1)}=Y_I{Z^I}}}
Using  the OPEs:
\eqn\grav{\eqalign{
\lambda_\alpha(z){\bar\mu}_\beta(w)
=-{\bar\mu}_\alpha(z)\lambda_\beta(w)
\sim{-{\delta_{\alpha\beta}}\over{z-w}}+...\cr
{\bar\lambda}_{{\dot{\alpha}}}(z)\mu_{{\dot{\beta}}}(w)
=-\mu_{{\dot{\alpha}}}(z){\bar\lambda}_{{\dot{\beta}}}(w)
\sim{{\delta_{{\dot{\alpha}}{\dot{\beta}}}}\over{z-w}}+...\cr
\psi_A(z)\psi_B(w)\sim{{\delta_{AB}}\over{z-w}}+...\cr
\bar\psi_A(z)\bar\psi_B(w)\sim{{\delta_{AB}}\over{z-w}}+...}}
one easily finds that,
 in the absence of the worldsheet instantons,
the conformal dimensions and the $GL(1)$ charges of $Y$ and $Z$
are respectively
\eqn\grav{\eqalign{h(Z^I)=0,h(Y_I)=1\cr
q(Z^I)=1,q(Y_I)=-1}}
In addition, as it is clear from the form of the action (1),
the $Y$ and $Z$  fields also can be attributed unit negative and 
positive ghost numbers
respectively:
\eqn\lowen{g(Z^I)=-g(Y_I)=1}
with the ghost number current having the same form as
 as the $GL(1)$ generator (2), so essentially $g=q$.
Given the stress tensor (2) there is no ghost anomaly,
however the  current does become anomalous when the worldsheet instantons are
present.
The $GL(1)$-neutral physical vertex operators of the theory
in the unintegrated $b-c$-picture are given by:
\eqn\grav{\eqalign{V_\phi=c{j^k}\phi_k(Z)\cr
V_f=cY_I{f^I}(Z),V_g=c\partial{Z^I}g_I(Z)}}
with the additional constraints on the space-time fields:
$\partial_I{f^I}=0,g_I{Z^I}=0$ and $j^k(z)$ being the dimension 1
currents from the current algebra, satisfying the usual OPE
(suppressing the conventional factor of $i$ before the structure constants):
\eqn\lowen{j_i(z)j_k(w)\sim{{k\delta_{ik}}\over{(z-w)^2}}
+{{f_{ikl}j^l(w)}\over{z-w}}+...}

Here $V_\phi$ corresponds to super Yang-Mills states while
$V_f$ and $V_g$ describe the conformal supergravity sector.
As the  operators are $GL(1)$-neutral, the $GL(1)$ charges
of $\phi_k,{f^I},g_I$ are $0,1$ and $-1$ respectively.
When the worldsheet instanton number $\int{d^2z}\epsilon^{ab}F_{ab}=0$,
these are the only massless operators, contributing to perturbative
correlation functions of the twistor-string theory.
In the presence of the worldsheet instantons, however, things change
 significantly. For the instanton number $d\neq{0}$ the expression
for the stress tensor (2) is deformed as
\eqn\lowen{T\rightarrow{T+{d\over2}\partial{J}}=T_c+Y_I\partial{Z^I}
+{d\over2}\partial{Y_IZ^I}}
implying the $GL(1)$ charge anomaly of the vacuum equal to $-d$.
To compensate for it, the total $GL(1)$-charge of the vertices entering
the correlator must be equal to $d$.
Furthermore, the conformal dimensions of the fields also get twisted
as
\eqn\grav{\eqalign{h(Z^I)\rightarrow{h}(Z^I)-{d\over2}=-{d\over2}\cr
h(Y_I)\rightarrow{h(Y_I)+{d\over2}}=1+{d\over2}}}
In general, the dimension of any space-time field $H_q(Z)$
carrying the $GL(1)$-charge $q$ is twisted according to
\eqn\lowen{h(H_q)\rightarrow{h(H_q)-{{qd}\over2}}}
Such a twist of course does not change the dimensions of the 
$GL(1)$-neutral operators (6) but if we are able to find
the charged $GL(1)$-operators commuting with the BRST-operator,
the anomalous dimension change must be taken into account.
Firstly, consider the simplest case $d=2$ and the operator
\eqn\lowen{V_2=cv\partial{Z^I}:j_ij_k:H^{ik}_I(Z)}
with the space-time field $H^{ik}_I$ carrying the $+1$
$GL(1)$-charge. The total charge of the operator is $+2$.
Note that, unlike the open  neutral operators
(6), the operator (11) carries both the $RP^{3|4}$ and the current algebra
indices.
Using the relations (9), (10) it is easy to see that the operator (11)
has conformal dimension zero.
Let us check its BRST-invariance.The BRST operator is given by {\mb}
\eqn\lowen{Q=\int{{dz}\over{2i\pi}}(cT-bc\partial{c}+v{Y_I{Z^I}}+
c{u\partial{v}})}
In the $d=2$ instanton background the operator (11) 
has conformal dimension zero and it is a primary field
provided that 
\eqn\grav{\eqalign{Tr(H)\equiv{H_I^{ik}}(Z)\delta_{ik}=0\cr 
f_{ikl}H_I^{kl}(Z)=0}}
 Therefore $V_2$ commutes with
the $cT-b{c}\partial{c}$ part of $Q$.Commutation with the remaining
the $\oint{c{u\partial{v}}}$ term of $Q_{brst}$ follows from
two terms follows from the OPEs of the fermions
 $c(z)c(w)\sim(z-w):\partial{c}c:(w)$
cancelling the simple pole from the $OPE$ of $u$ and $v$,
so the full OPE of $\oint{cu}\partial{v}$ and $V_2$ is non-singular.
Finally,using  $v(z)v(w)\sim(z-w):\partial{v}{v}(w)$,
 the commutation with the ${\oint}vJ_{GL(1)}$ term gives
\eqn\lowen{
\lbrace{\oint}{{dz}\over{2i\pi}}:vY_I{Z}^I:(z),V_2(w)\rbrace
=cv\partial{v}Z^IH_I^{ik}(Z(w)):j_ij_k:(w)=0}
provided that we impose the constraint on $H$:
\eqn\lowen{Z^IH_I^{ik}(z)=0}
Now let us consider the question of the BRST non-triviality of
$V_2$. The potential BRST-triviality threat comes from the operator
$W_2=c\partial{Z^I}j_ij_kH_I^{ik}(Z)$.
Using the BRST-invariance condition (15), one easily computes
\eqn\lowen{\lbrace{Q_{brst},W_2\rbrace}=
V_2+cv\partial{Z^I}:j_ij_j:Z^J\partial_J{H_I^{ij}}(Z)}
which implies that the $V_2$ operator is BRST notrivial, provided that
\eqn\lowen{Z^J\partial_J{H_I^{ik}(Z)}\neq{0}}.
Equation (17) implies an important subtlety.
Namely, at the first glance, it is not evident that, even if (17)
is satisfied, this guarantees the non-triviality of the $V_2$-operator.
Indeed, since the $H$-field (17) has a $GL(1)$ charge $+1$ and, generically,
$Z^J\partial_j$ plays the role of the $GL(1)$-generator, naively
one can write
\eqn\lowen{Z^J\partial_JH^{ik}_J(Z)=H^{ik}_J(Z)}
and then (16),(17) would seem to imply $V_2=\lbrace{Q_{brst},V_2}\rbrace$.
More generally, for any $H$-field of the $GL(1)$ charge $Q$ one would
have
\eqn\lowen{Z^J\partial_J{H(Z)}=qH(z)}
In fact, however, the relations (18), and (19) are not always true -
and whenever they are not fulfilled, $V_2$ is non-trivial.
Let us address this question in more details.
Whether (18) and (19) are true, depends completely
on the form the $H$-fields depend on the homogeneous $Z$-coordinates.
For instance, (18) and (19) are always true
if H is some homodgeneous polynomial of $Z^I$,
e.g. $H^{I_1...I_k}=Z^{I_1}...Z^{I_k}$, as it is easy to check.
On the other hand, (18) and (19) are not satisfied if H contains
any logarithmic dependence, e.g.
\eqn\grav{\eqalign{H^{I_1...I_q}(Z)=Z^{I_1}...{Z^{I_q}}
log({\prod_{j=1}^n{Z^{I_j}}})\cr
Z^J\partial_J{H^{I_1...I_q}(Z)}=qH^{I_1...I_q}(Z)+
nZ^{I_1}...Z^{I_q}\neq{H^{I_1...I_q}}}}
At this point, the following remark should be made.
Naively, the introduction of the logarithms in $H$
would not seem to be a right idea, since
 under the GL(1) transformation $Z\rightarrow{tZ}$
the logarithm is translated by $t$, seeming
to  imply that it doesn't have associated well-defined 
$GL(1)$-charge and therefore we cannot talk definitely
about $GL(1)$ charge and hence the definite
 conformal dimension of the H-field in (20).
However we will show that, in the sense explained below,
the logarithm of $Z$  $can$ in fact be regarded as a
field of $GL(1)$ charge zero and conformal dimension zero in any
worldsheet instanton background.
Namely, consider the instanton background $d=2N$ and the
associated stress-energy tensor for the $YZ$-system:
$$T(z)=Y_I\partial{Z^I}+N\partial(Y_IZ^I)=
(N+1)Y_I\partial{Z^I}+NZ^I\partial{Y^I}$$
Then the OPE of this tensor with $log{Z^J}$
gives
\eqn\grav{\eqalign{T(z)log{Z^J(w)}=(N+1){{\partial{Z^J(w)}}\over{Z^J}}
+NZ^J(w)\partial_w{1\over{Z^J(w)}}\cr=
(N+1)\partial(log{Z^J(w)})-N\partial(log{Z^J(w)})=
\partial(log{Z^J(w)})}}
implying that $log{Z^J}$ is a primary field of dimension 0
in any $d=2N$ worldsheet instanton background.
But then the equation (10) which defines the correspondence  between the
$GL(1)$ charges and anomalous conformal dimensions in the instanton 
backgrounds, implies that the $GL(1)$ charge of the $log{Z^J}$ field
must also be set to zero.
This means that inclusion of the logarithms in the space-time $H$-fields
(17),(20) does not change their $GL(1)$ charges or conformal dimensions
and at the same time protects their BRST non-triviality according to (17).
Thus the coordinate dependence of the
target space fields of the physical $GL(1)$ charged
states always needs to include the logarithms.

The conditions (13),(15), and (17) on the $H$-field insure that the 
$V_2$-operator is physical, i.e. BRST-invariant and non-trivial.

It is important that the explicit dependence of the $V_2$-operator
on the $GL(1)$ ghost $v$ field protects it from having a simple pole 
due to the OPE with $J=Y_IZ^I$ because of its nonzero $GL(1)$ charge
(provided that the constraint (15) is satisfied.
The ghost number anomaly of the $u-v$ system is equal to $-1$,
as the $u$ and $v$ ghosts have dimensions 1 and 0 and therefore 
can be bosonozed as $u=e^{-\chi},v=e^\chi$ with $\chi$ being a free
field with stress-energy tensor 
$T_\chi={1\over2}((\partial\chi)^2+\partial^2\chi)$.
As the $Y-Z$ ghost anomaly for tree level correlator is equal to $-2$
 for $d=2$, and the $GL$-neutral operators do not depend on $u$ and $v$,
it appears that, unless some consistent
picture-changing formalism is introduced,
 the only non-vanishing correlators 
(with the ghost anomaly cancelled)
would involve not more than one $V_2$-operator
with arbitrary number of $GL(1)$-invariant vertices.
The form of dimension 0 BRST-invariant picture-changing operator
lowering the $u-v$ ghost number by one unit can be found relatively easily
 from the integration over the moduli of the $A_{z,{\bar{z}}}$ 
$GL(1)$-connection
present in the worldsheet action (1). The integration over moduli 
of the worldsheet $GL(1)$ gauge field leads, as usual
(along with the standard condidion that the basis of the moduli
is orthogonal to the gauge slice)
to the insertion of the picture-changing operators
inside the correlation functions, having the form
\eqn\lowen{\Gamma(z)=u\delta(J_{GL(1)})=u\delta(Y_IZ^I)}
Here the $u$ ghost has the $+1$ dimension and the 
delta-function of the $GL(1)$-current 
is of the dimension $-1$, so the total dimension of $\Gamma$ is zero.
The BRST-invariance of $\Gamma$ can be easily checked by using
\eqn\grav{\eqalign{\lbrace{Q_{brst}},u\rbrace=Y_IZ^I\cr
\lbrack{Q_{brst},Y_IZ^I}\rbrack=0\cr
:Y_IZ^I\delta(Y_IZ^I):\sim{0}}}
Note that the  $R$-operator,
introduced in the paper by Berkovits and Motl {\mb}
and which $increases$ the $u-v$ ghost number by 1:
$R=v\delta(r(z)-1)+c\partial{r}\delta(r(z)-1)$
where $r$ is the scale factor for the $Z$-field
should in fact be regarded as the  $inverse$ rather than
the direct picture-changing  operator.
Note that unlike the case of $\Gamma$, it is not yet clear how
the $R$-operator can be derived from the first principles, i.e. from the 
the integration over moduli.
Using the appropriate insertions of  $\Gamma$ and $R$-operators
 in the correlators,
one can  in principle consider the amplitudes with arbitrary number
of the $V_2$-vertices.
 In this paper, however, we shall only consider the correlators
not requiring any picture-changing insertions, for the sake of simplicity.

As an example, consider the 3-point correlation function of the $V_2$-operator
with 2 gluon vertices at $d=2$. Simple computation gives
\eqn\grav{\eqalign{
<cv\partial{Z^I}:j_ij_k:(z_1){H^{I}_{ij}}(Z)c{j^l}(z_2)\phi_l(Z)
c{j^m}\phi_m(z_3)>\cr=\int{d^4x}d^8\theta\int_{D_{x,\theta}}dZ^I
{{H^{ik}_I}(Z)\phi_i\phi_k{}}}}
where the degree 1 curve $D_{x,\theta}$ is defined through
$\mu^{{\dot{\alpha}}}=x^{\alpha{\dot{\alpha}}}\lambda_\alpha,
\psi^A=\theta^{A\alpha}\lambda_\alpha$
with $\lambda$ being the homogeneous  coordinates and
$(x,\theta)$ the moduli of the curve.
Let us now turn to the general case
of $d=2N$. Repeating the above arguments, it is easy to see that
the spectrum of massless physical GL(1)-charged states is given by the 
vertex operators of the form:
\eqn\grav{\eqalign{V_{2N}(w)=cv\partial{Z^{I_1}}...\partial{Z^{I_P}}{\lbrack}
H_{I_1...I_P}^{i_1...
i_{m(P,N)}}(Z(w))\rbrack_{{q_{GL(1)}(H)=2N-P}}:j_{i_1}....j_{i_{m(N,P)}}
:(w)\cr
m(N,P)=2N^2-P+1;{P\leq{2N}}}}

The $I_k$ are the target space indices of $RP^{3|4}$ while
the $i_k$ indices  are related to the current algebra (7). 
Each of these operators has the total $GL(1)$ charge $q=2N$
so that one insertion of $V_{2N}$ in correlation functuions compensates for
the $Y-Z$ ghost anomaly. The space-time $H$-fields are allowed to have
$GL(1)$ charges equal to $q_{GL(1)}=2N-P,P=0,...2N$.
As it is easy to check, the number $m(N,P)$ of the 
dimension 1 current insertions 
insures that the total conformal dimension of the operators (25) 
is equal to zero. 
They are symmetric in both the target space and the current algebra indices.
The BRST invariance and non-triviality conditions 
are derived  completely similarly to the case of $V_2$.
The BRST-invariance of $V_{2N}$
requires
\eqn\grav{\eqalign{Z^{I_1}H_{I_1...I_P}^{i_1...i_{m(N,P)}}(Z)=0\cr
H_{I_1...I_P}^{i_1...i_k...i_l...i_{m(N,P)}}(Z)\delta_{i_ki_l}=0\cr
f_{ii_ki_l}H_{I_1...I_P}^{i_1...i_k...i_l...i_{m(N,P)}}(Z)=0
}}
(as $H$ is symmetric in the current indices,
the second and the third identities hold for any pair
of $i_k$ and $i_l$).
The BRST non-triviality of $V_{2N}$ requires
\eqn\lowen{Z^J\partial_J{H_{I_1...I_P}^{i_1...i_{m(N,P)}}}(Z){\neq}0}
As previously, the last equation also must be supplemented with the condition
that the $Z$-dependence of the $H$-field must contain  logarithms.

Now we are prepared to explore some properties of the $V_{2N}$-operators.
As all of these operators contain the $v(w)$ ghost field
(protecting their BRST-invariance), the conditions for cancelling
the $u-v$ and $Y-Z$ ghost anomalies altogether require that not
more than one $V_{2N}$ insertion can be present in the correlations
(all other vertices must be $GL(1)$-neutral).
For this reason, in this paper  we will be interested in the
conformal beta-functions of the $H$-fields stemming from the 
singularities of the OPE between $V_{2N}$-operators and
$GL(1)$-neutral open and closed string vertices
$j^i\phi_i,f_IY^I$ and $g_I{\partial}Z^I$.
First of all, it is easy to see that the OPE of
$V_{2N}$ and $g_I(Z)\partial{Z^I}$ as non-singular and therefore 
does not contribute to the beta-function of the $H$-field
(since the latter vertex contains no $Y$ or $j$ variables)
Next, the OPE simple pole singularity between $V_{2N}$ and
$f_IY^I$ stems from two products:
\eqn\grav{\eqalign{Y^I(z)H^{I_1...I_P}_{i_1...i_{m(P,N)}}
(Z(w))\sim{1\over{z-w}}\partial_IH^{I_1...I_P}_{i_1...i_{m(P,N)}}(Z(w))
\cr
Y_I(z)\partial{Z^J}(w)\sim{{\delta_I^J}\over{(z-w)^2}}}}
This leads to the OPE
\eqn\grav{\eqalign{cv\partial{Z^{I_1}}...\partial{Z^{I_P}}
:j_{i_1}...j_{i_{m(P,N)}}:H_{I_1...I_P}^{i_1...i_{m(P,N)}}(Z(z))
f_IY^I(w)\cr
\sim{1\over{z-w}}cv\partial{Z^{I_1}}...\partial{Z^{I_P}}
:j_{i_1}...j_{i_{m(P,N)}}:(z)\times\lbrace
f^I\partial_I{H^{i_1..i_{m(P,N)}}_{I_1...I_P}}+P\partial_{I_1}
f^JH_{JI_2...I_P}^{i_1...i_{m(P,N)}}\rbrace}}
In other words, the OPE between the charged and the neutral operators
has a simple form 
\eqn\lowen{V_{2N}V_f\sim{1\over{z-w}}CV_{2N}} (where C are
the relevant structure constants), as one should  have expected.
This means that this OPE singularity leads 
in the product of the charged and neutral states leads
to the renormalization
of the $H$-field, but not the $f^I$-field of the neutral operator.
The above OPE allows us to immediately write the 
renormalization group equation for the $H$-field:
\eqn\lowen{ \Lambda{{d}\over{d\Lambda}}{H^{i_1..i_{m(P,N)}}_{I_1...I_P}}
=-{f^I\partial_I}{H^{i_1..i_{m(P,N)}}_{I_1...I_P}}-P\partial_{I_1}f^J
{H^{i_1..i_{m(P,N)}}_{JI_2...I_P}.}}
Let us now turn to the interactions of $V_{2N}$ with the open string
 neutral $V_\phi$-vertices (6) and the related contributions to the
RG flows of the $H$-field. While it is quite obvious  that schematically
the OPEs of $V_{2N}$ and $V_\phi$  have the  structure  similar to 
(30), few important subtleties appear due to the anomalous term
in the Kac-Moody algebra (7). Contributions to the RG flow of the $H$-field
originate from both of the terms in (7). Simple calculation gives:
\eqn\grav{\eqalign{cv{\partial}Z^{I_1}...\partial{Z^{I_P}}
:j_{i_1}...j_{i_{m(P,N)}}:
{H^{i_1..i_{m(P,N)}}_{I_1...I_P}}(Z(z)):j^k(w)\phi_k(Z(w)):
\cr\sim{1\over{z-w}}cv{\partial}Z^{I_1}...\partial{Z^{I_P}}
{H^{i_1..i_{m(P,N)}}_{I_1...I_P}}(Z(z))
:{j_{i_1}...j_{i_{m(P,N)-1}}}:(z)\cr\times\lbrace{-\kappa\delta_{ki_{m(P,N)}}
\partial_{I_{P+1}}\phi_k\partial{Z^{I_{P+1}}}+
f_{i_{m(P,N)}kl}{\phi_k}j_{l}(w)}\rbrace
}}
The OPE singularity involving the structure constants of the 
current algebra (7) leads to logarithmic divergence in the partition function 
which can be removed by the RG flow of the 
${H^{i_1..i_{m(P,N)}}_{I_1...I_P}}$-field
 the term given by $\sim{f_{i_1kl}}\phi_k{H^{li_2..i_{m(P,N)}}_{I_1...I_P}}$
(as previously, the symmetrization over the $i_k$ current indices is implied)
On the contrary, the divergence caused by the anomalous
term of the current algebra, cannot be removed by the flow of this
$H$-field, as this term involves the extra $\partial{Z}$-factor.
Therefore to remove this divergence we need to renormalize
$another$ $H$-field with the $GL(1)$-charge differing by 1 unit:
\eqn\grav{\eqalign{
\lbrack{H^{i_1..i_{m(P+1,N)}}_{I_1...I_{P+1}}}\rbrack_{q=2N-P-1}
{\rightarrow}
\lbrack{H^{i_1..i_{m(P+1,N)}}_{I_1...I_{P+1}}}\rbrack_{q=2N-P-1}\cr+
{\kappa}log\Lambda\lbrack{{H^{i_1..i_{m(P+1,N)}i_{m(P,N)}}_{I_1...I_P}}}
\rbrack_{{q=2N-P}}
\partial_{I_{P+1}}\phi^{i_{m(P,N)}}}}
(recall that $m(P,N)=2N^2-P+1$)
Here the $q=2N-P-1$ is the $GL(1)$ charge of the
 space-time $H$-field generated by the vertex
operator $${cv}\partial{Z^{I_1}}...\partial{Z^{I^{P+1}}}
:j_{i_1}...j_{i_{m(P+1,M)}}:{H^{i_1..i_{m(P+1,N)}}_{I_1...I_{P+1}}}$$
containing the extra $\partial{Z}$ factor that we needed.
Summarizing the above, the total RG flow of the 
$H$-field due to the OPE of $V_{2N}$ and the open string
vertex $V_\phi$ is given by
\eqn\grav{\eqalign{\Lambda{{d}\over{d\Lambda}}
\lbrack{H^{i_1..i_{m(P,N)}}_{I_1...I_P}}\rbrack_{q=2N-P}
=-f_{i_1kl}{\phi^k}\lbrack{H^{li_2..i_{m(P,N)}}_{I_1...I_P}}\rbrack_{q=2N-P}
\cr+\kappa\partial_{I_P}\phi^{i_m(P-1,N)}
\lbrack{H^{i_1..i_{m(P,N)}i_{m(P-1,N)}}_{I_1...I_{P-1}}}\rbrack_{q=2N-P+1}
}}
One important consequence of this equation is that
the  anomaly in the Kac-Moody algebra leads to the non-conservation
of the $GL(1)$-charge in the interactions involving the $V_{2N}$-vertices
(as the RG flow of the $H$-field with $q=2N-P$ involves 
another target space field with $q=2N-P+1$).
In the next section we shall attempt to discuss
the physical reasons behind such a non-conservation.
The general idea is that the $V_{2N}$-operators discussed in this paper
describe the collective coordinates of the solitons of 
the conformal supergravity and super Yang-Mills theory.
 The charge non-corservation is thus
the non-perturbative effect, related to the interaction between the solitons.

\centerline {\bf Conclusion and Discussion}

In this letter we analyzed the properties of the $GL(1)$-charged vertex 
operators in twistor string theory, emerging in the presence of the
worldsheet instantons. 
But the main question remains - what physical degrees of freedom
these operators correspond to? Once we know that the 
perturbative massless spectrum of the twistor string theory is limited
to the neutral open and closed string states 
(corresponding to $N=4$ super Yang-Mills theory and the conformal
supergravity), then what is the role played by the tower of the charged
massless states?
Below we shall attempt to present some arguments
 that these states are related to the 
non-perturbative sector of the twistor-string theory, possibly corresponding
to the collective coordinates of non-perturbative solitonic 
(or brane-like) solutions
of the conformal supergravity theory.
To justify such a conjecture, let us
make an excursion to the NSR superstring theory, in order to point out
 some important analogies with 
twistorial strings.
While the conventional perturbative vertices
in twistor string theory are GL-neutral (which usually is the consequence of 
their BRST properties)
 the BRST-invariance of the charged vertices
is insured by their explicit dependence on the $v$ ghost field,i.e.
these operators are essentially mixed with $GL(1)$-ghosts.
Moreover, the $GL(1)$-charge is not conserved
in the interactions involving the $V_{2N}$-vertices, as has been demonstrated
 by the analysis of their renormalization group flows (34). At the same time, 
the operator algebra involving the products of $V_{2N}$-operators
with $GL(1)$-neutral states still has a remarkably simple structure (30).
In NSR superstring theory there  exists a set of physical operators
with remarkably similar properties.
These operators are also non-invariant under local worldsheet supersymmetry,
however their coupling with superconformal ghosts 
(which cannot be removed by picture-changing transformations
and therefore is not an artefact of a picture choice)
insures their physical BRST properties. These operators exist in both
 open and closed string sectors and are related to non-perturbative 
dynamics of branes. Examples of such  open-string operators 
are given  by the five-form states
\eqn\grav{\eqalign{V_5^{(-3)open}=H^{m_1...m^5}(k)
\oint{{dz\over{2i\pi}}}e^{-3\phi}\psi_{m_1}...\psi_{m_5}e^{ikX}\cr
V_5^{(+1)closed}=H^{m_1...m^5}(k)
\oint{{dz\over{2i\pi}}}e^{\phi}\psi_{m_1}...\psi_{m_5}e^{ikX} 
+ghosts\cr
k^{\lbrack{m_1}}H^{m_2...m_6\rbrack}\neq{0}}}
This five-form state exists at ghost pictures above $+1$ or below $-3$
but, say, not at the picture zero, so its mixing with superconformal ghosts
is essential.
The zero-momentum part of the picture {-3} appears as a five-form extension
in the picture-twisted space-time superalgebra of NSR superstring theory:
\eqn\grav{\eqalign{\lbrace{T_\alpha,T_\beta}\rbrace=\gamma^m_{\alpha\beta}P_m
+\gamma^{mn}_{\alpha\beta}Z_{mn}+\gamma^{m_1...m_5}_{\alpha\beta}Z_{m_1...m_5}
\cr
T_\alpha=\oint{{dz}\over{2i\pi}}
(e^{{-3\over2}\phi}+e^{-\phi\over2})\Sigma_\alpha\cr
P_m=\oint{{{dz}\over{2i\pi}}}e^{-\phi}\psi_m;Z_{mn}=\oint{{dz}\over{2i\pi}}
e^{-2\phi}\psi_m\psi_n\cr
Z_{m_1...m_5}=\oint{{dz}\over{2i\pi}}e^{-3\phi}\psi_{m_1}...\psi_{m_5}
}}
where $\Sigma$ is the spin operator for matter fields, $\phi$ is the
bosonized superconformal ghost and $\psi$ are the NSR worldsheet fermions.
The superalgebra (36) is isomorphic to the dimensionally reduced
M-theory algebra, or the SUSY algebra of $d=11$ supergravity {\town}.
It has been pointed out  {\town} that the central $p$-form terms in the 
SUSY algebra  $always$ correspond to the presence
of extended objects (p-branes) in the theory.
Thus one shall attempt to relate the 2-form and the 5-form
vertices (36) to topological  charges
 of the M-branes, i.e. purely perturbative physical vertices account
for the dynamics and topology of essentially non-perturbative objects
(e.g. the non-perturbative solutions of $d=11$ supergravity).
Moreover, many analogues of these states exist in the closed string sector,
 some of them can  be obtained e.g.
by taking the product of the left 5-form and the photon-type right-moving part:
\eqn\grav{\eqalign{V_5^{closed}=H^{m_1...m_6}\int{d^2z}(e^{\phi-\bar\phi}+
e^{-3\phi-\bar\phi})
\psi_{m_1}..\psi_{m_5}\bar\psi_{m_6}e^{ikX}+ghosts\cr
k_{m_6}H^{m_1...m_5m_6}=0\cr
k^{{\lbrack}m_7}H^{m_1...m_5\rbrack{m_6}}\neq{0}
}}
Here the 6-tensor $H$ is antisymmetric in the first 5 indices. Using the BRST
conditions (31) on $H$, one can eliminate half of its independent components.
Then for each fixed orientation $m_1...m_6$ of H in the space-time the
 constraints
(37) imply that there are six independent degrees of freedom
$\lambda^t,t=1...6$ per orientation and 
the momentum $k$ must me orthogonal to the
$m_1...m_6$ directions 
(the exact relation between $\lambda$ and $H$ is given in {\self})
 Using the above conditions one can calculate
the effective action for the fixed orientation of $H$ to be given by
\eqn\grav{\eqalign{S_{eff}\sim{\int{d^4x}}e^{-\varphi}{\sqrt{det(\eta_{ab}+
\partial_a\lambda^t\partial_b\lambda^t)}}\cr
a,b=0..3}}
i.e. the DBI effective action for D3-brane.
On the other hand, the open string $V_5$ operators
are responsible for the RR-coupling terms and topological charges of 
various brane configurations.
Finally, the OPE algebra of the $V_5$ operators 
(both in the open and the closed string cases) with usual 
perturbative NSR vertices
 has a structure totally similar to (30) of the twistor string case,
i.e. {\self}
\eqn\lowen{V_5(z){V}(w)\sim{1\over{z-w}}CV_5({{z+w}\over2})}
where C are the structure constants and V is either 
a photon, a dilaton , a graviton or an axion.
We have brought up the recollection of the NSR scenario
to emphasize our claim that the $V_{2N}$-vertices discussed in this letter 
have properties very similar to (35), (36) and (39)
correspond to the non-perturbative solitonic objects that should appear in the 
conformal supergravity. At present, such solitonic
(or brane-type) solutions  have not yet been 
discussed in the context of  conformal supergravity and it 
would be an interesting problem to explore. We hope to elaborate on it
in the future works. The existence of physical $V_{2N}$-operators
seems to clearly predict the possibility of such non-perturbative solutions
in the worldsheet instanton backgrounds.
Once the conformal supergravity solitons are found,
it would be important to calculate the effective actions
governing their dynamics, which should be somewhat of the DBI-type.
In principle these actions can be obtaind straightforwardly from the 
correlations of the $V_{2N}$-operators. However, such a computation would 
require a well-defined and complete formalism 
of the  $GL(1)$-ghost picture changing
(both the direct and the inverse) developed for twistor string theory,
in the manner it is available in the NSR case.
Unfortunately, at present we do not yet have such a formalism, and
its construction would clearly need some further efforts.

\centerline{\bf Acknowledgements}

I am grateful to Wafic Sabra for useful comments and 
discussions on the subject. I'm also grateful to N.Berkovits for his useful
remarks on the non-triviality of the $V_{2N}$-vertices.

\listrefs
\end